# ON KLJN-BASED SECURE KEY DISTRIBUTION IN VEHICULAR COMMUNICATION NETWORKS

X. Cao [1,3], Y. Saez [1,+], G. Pesti [2], L.B. Kish [1]

[1] *Department of Electrical Engineering, Texas A&M University, College Station, TX 77843-3128, USA*
yessica.saez@tamu.edu; laszlo.kish@ece.tamu.edu

[2] *Texas A&M Transportation Institute, Texas A&M University, College Station, TX 77843-3135, USA*
g-pesti@tamu.edu

[3] *College of Automotive Engineering, Jilin University, Changchun, Jilin 130025, China*
caoxiaolin@jlu.edu.cn

In a former paper [*Fluct. Noise Lett.*, **13** (2014) 1450020] we introduced a vehicular communication system with unconditionally secure key exchange based on the Kirchhoff-Law-Johnson-Noise (KLJN) key distribution scheme. In this paper, we address the secure KLJN key donation to vehicles. This KLJN key donation solution is performed lane-by-lane by using roadside key provider equipment embedded in the pavement. A method to compute the lifetime of the KLJN key is also given. This key lifetime depends on the car density and gives an upper limit of the lifetime of the KLJN key for vehicular communication networks.

## 1. Introduction

After more than 100 years of development on modern vehicle technology, we are on our way to smarter cars and much more intelligent transportation systems. Nowadays, people pay more attention to safety and comfort in vehicles, rather than traditional traction ability, fuel cost, handling, and stability. Therefore, it is believed that safety

[+] Corresponding Author, *yessica.saez@tamu.edu*

and mobility information such as road and traffic information (e.g. emergency braking, vehicle collision, congestion, toll collection, etc.), weather forecast warnings (e.g. water or ice on the pavement), and local services (e.g. route maps, gas or restaurant locations, etc.)[1–4] should be collected and provided to drivers and passengers in the vehicle.

### 1.1. *Vehicular Communication Networks*

Vehicular communication networks have become a reasonable intelligent transportation solution that can satisfy these demands effectively. A typical vehicular communication network is shown in Fig. 1[1, 5–9].

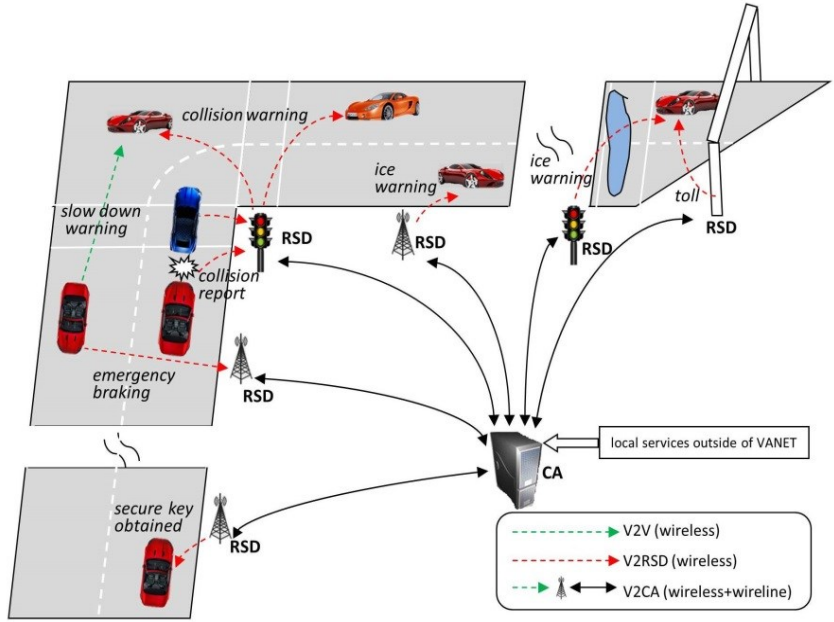

Fig. 1. A typical vehicular communication network. Three basic nodes are encountered in this type of network: Vehicles, Roadside Devices (RSDs) and Certification Authorities (CAs). The types of communication within vehicular communication networks include [1, 5–9]: Vehicle-to-Vehicle (V2V), Vehicle-to-Roadside-Device (V2RSD), and Vehicle-to-Certification-Authority (V2CA).

As summarized in previous publications [1, 5–9], Vehicles, Roadside Devices (RSDs), and Certification Authorities (CAs) are the three basic nodes in most of vehicular communication networks. Vehicles are mobile terminal nodes that are in charge of collecting road and traffic information, reporting events to the CAs through the RSDs, and exchanging warnings with nearby vehicles. The RSDs are intermediate nodes in charge of transferring messages between vehicles and CAs in two-ways. The CAs are the host nodes that manage information related to vehicles. These nodes also generate secure keys and provide certifications for all vehicles in the network, control message exchanges of the whole network, and distribute local information obtained outside the local vehicular communication network. Accordingly, the types of communication within vehicular communication networks include [1, 5–9]: Vehicle-to-Vehicle (V2V), Vehicle-to-Roadside-Device (V2RSD), and Vehicle-to-Certification-Authority (V2CA). Communications within vehicular communication networks raise concerns for security and privacy. For example, the identity of vehicles, emergency braking, and vehicle collision warnings among vehicles must be transmitted securely to avoid malicious activities. The private financial information used in toll collection when cars pass by RSDs also needs to be protected.

In order to solve these fundamental security-related issues for promising vehicular communication network applications, several security protocols have been proposed by different researchers. In [10–11], the authors proposed a security infrastructure that is based on public key infrastructure (PKI). Later, other solutions based on PKI were proposed [2, 4, 12, 13]. The authors of [2] provided a "lightweight" authenticated key scheme that integrates blind signature techniques for V2V and V2RSD communications. In [4], the authors presented an approach that combines the traditional PKI and identity-based public key cryptography for vehicular communication networks. In [12], a secure scheme with session keys (pairwise and group keys) used in non-safety-related applications (e.g. "chatting in platoon") was designed. In [13], temporary anonymous certified keys (TACKs) were constructed, and a key management scheme based on TACKs was proposed for vehicular communication networks. Besides PKI, group signatures are another important category of proposed security methods. Based on the strong Diffie-Hellmanand linear assumptions, the authors of [14] introduced the under-200 bytes group signature scheme that has a similar security level to the RSA (Rivest, Shamir, and Adleman public-key cryptosystem) signature of the same length. A group signature-based protocol using tamper-resistance devices and a probabilistic signature verification scheme was proposed in [3]. In [15], the authors constructed an identity-based batch verification scheme for V2RSD communication in vehicular communication networks. In [16], a software-based roadside unit-aided messages authentication protocol for V2V communications was proposed. In addition, a software-based solution that uses secure and privacy enhancing communication schemes for vehicular sensor networks was provided in [17].

Most of the above security schemes or protocols are constructed based on software encryption mechanisms. The security on these software-based methods is based on the premise that the eavesdroppers have *limited computational power*. Thus, these security schemes offer just a *computationally conditional security* [18–22]. Moreover, these architectures focus their attention on V2V or V2RSD communications and although there is significant information transmitted in the Road-side-Device-to-Certification-Authority (RSD2CA) communication [9], it is very rare to find works related to securing this particular communication channel.

In [9], a novel unconditionally secure vehicular communication architecture that utilizes the KLJN key distribution scheme was proposed. In this architecture, a new node called the Roadside-Key-Provider (RSKP) was introduced to provide the cars with KLJN keys. Based on this work, we discuss the KLJN-based secure key generation, donation, and lifetime in vehicular communication networks. The remainder of this paper is organized as follows. In section 2.1, we discuss the key generation process in vehicular networks by describing the message exchange and specifying the different network lines used during the process. In section 2.1, we propose a lane-by-lane KLJN key donation solution for the vehicular communication architecture with unconditionally secure key exchange proposed in [9]. An upper limit for the KLJN key lifetime in vehicular communication networks is computed in section 2.3. The results are demonstrated with practical considerations. Section 3 concludes the paper.

## 1.2. *On the KLJN key exchange*

The illustration of the ideal KLJN key exchange scheme is shown in Fig. 2 [19, 21–24].

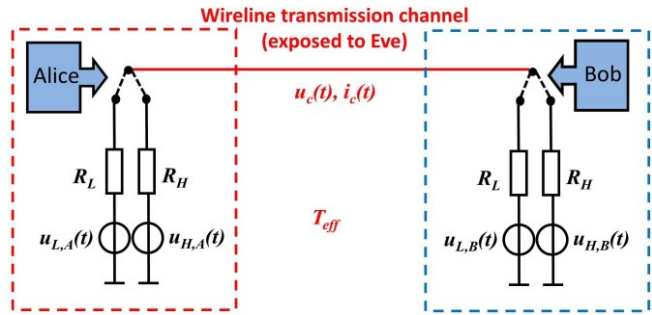

Fig. 2. Illustration of the ideal KLJN key distribution scheme. There is a switch and identical pairs of resistors ($R_L$ and $R_H$, $R_L \neq R_H$) on each communicator (referred to as Alice and Bob) side, where $R_L$ represents the low, $L$ bits , $R_H$ represents the high, $H$ bits; $u_{L,A}(t)$, $u_{H,A}(t)$ and $u_{L,B}(t)$, $u_{H,B}(t)$ are the thermal (Johnson) noise voltages (Gaussian noise voltage generators) at temperature $T_{eff}$ of $R_L$ and $R_H$ of Alice and Bob, respectively, $u_c(t)$ is the channel noise voltage, and $i_c(t)$ is the noise current in the wire.

The Kirchhoff-law-Johnson-noise (KLJN) secure key exchange scheme was proposed in 2005 [19] as a statistical/physical competitor to quantum key distribution (QKD) [20]. The KLJN scheme provides unconditional security based on the Kirchhoff's loop law of quasi-static electrodynamics and the fluctuation-dissipation theorem [19, 21–24]; for a general security proof, see [25]. Several potential applications have been proposed such as: classical networks [26], smart power grids [27], and secure computers, algorithms, and hardware [28].

In this ideal KLJN scheme, the two communicating parties, Alice and Bob, communicate via a wireline channel. There is a switch and identical pairs of resistors ($R_L$ and $R_H$, $R_L \neq R_H$) on each communicator side, where $R_L$ represents the low, *L* bit and $R_H$ represents the high, *H* bit. The Gaussian noise voltage generators $u_{L,A}(t)$, $u_{H,A}(t)$ and $u_{L,B}(t)$, $u_{H,B}(t)$ represent the enhanced thermal (Johnson) noise voltages at temperature $T_{eff}$ of $R_L$ and $R_H$ of Alice and Bob, respectively, while the $u_c(t)$ is the channel noise voltage, and the $i_c(t)$ is the noise current in the wire. At the beginning of the bit sharing period, both Alice and Bob *randomly* choose one of the resistors ($R_L$ or $R_H$) and the corresponding Gaussian noise voltage generator ($u_{L,A}(t)$ or $u_{H,A}(t)$, $u_{L,B}(t)$ or $u_{H,B}(t)$, respectively) and connect them to the wireline. The possible permutations of the resistors connected to the channel will be: *LL, LH, HL*, and *HH*. In the cases of *LL* or *HH*, the location of the resistors and the exchanged bits are publicly known, thus these bits are discarded [19]. On the other hand, *LH* and *HL* are secure bit exchange situations, because Eve cannot differentiate between the situations *LH* and *HL*.

The security of the ideal KLJN scheme is based on the second law of thermodynamics [19, 21–24], that is the difficulty to crack the ideal KLJN system is similar to that of to build a perpetual-motion machine of the second kind. In addition, the KLJN system is robust and not sensitive to vibrations [24], and is easy to be integrated on chips [28]. Based on the above core scheme, some advanced schemes were proposed to enhance the speed of the KLJN system [23].

In order to protect against active (invasive) attacks (and also against passive attacks on non-ideal systems), the KLJN system continuously monitors or measures the instantaneous current and voltage at the two ends of the line [22]. These measurements are compared via an authenticated public channel. Therefore, any intruder causing changes in the circuitry, and thus affecting the instantaneous measurements, will cause an alarm to go off and Alice and Bob will discover the intrusion. This defense mechanism is illustrated in Fig. 3.

*On KLJN-based secure key distribution in vehicular communication networks*

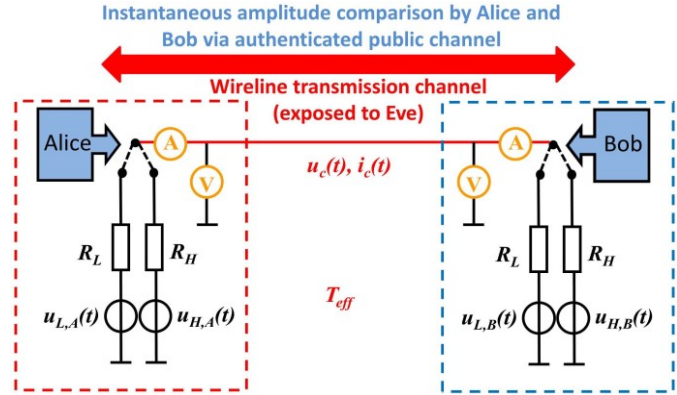

Fig. 3. KLJN system protection against invasive (active) attacks. Alice and Bob measure the instantaneous channel voltage and current amplitudes and compare them via an authenticated public channel. Alice and Bob learn all the information Eve can have. Additional elements to prevent hacking—such as line filters, line capacitance killer arrangement, etc.—are not shown. The notation is the same as in Fig. 2.

It is important to note that the instantaneous current and voltage data contain all the information related to the key that Eve could have. Thus, it is impossible for Eve to extract key information without letting the system know of such activity. In consequence, Alice and Bob can decide whether or not to discard the compromised bits according to a previously agreed maximum allowed level of information leak toward Eve [22].

According to the working principle of the KLJN scheme, the secure bit exchange takes place when the resistor states of the two communicators (*i.e.,* the CA and RSD and/or RSKP in vehicular communication networks) are different, *i.e., LH* or *HL*. This is indicated by an intermediate level of the mean-square noise voltage ($u_{msn}$) on the line, or that of the noise current in the wire [19]. This concept is shown in Fig. 4.

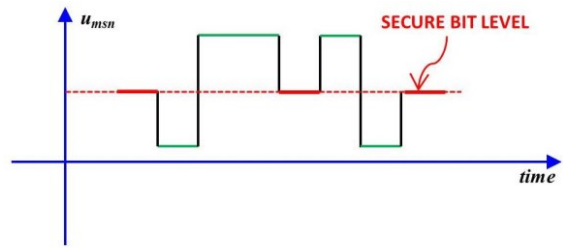

Fig. 4. The secure bit generation in the KLJN scheme. The intermediate mean-square noise level represent the bit situations *LH* or *HL*, that is when a secure bit exchange takes place.

It is important to mention that the two communication parties must previously and publicly agree on which one of them will invert the exchanged bit to have identical keys at the two ends.

## 2. KLJN Secure Key in Vehicular Communication Networks

Based on the working principle of the KLJN scheme [19] and the vehicular communication network model with unconditional secure key exchange proposed in [9], we discuss the generation, donation, and lifetime of the KLJN secure key in vehicular communication networks.

### 2.1. *KLJN key generation in vehicular communication networks*

According to the vehicular communication network model with unconditional secure key distribution proposed in [9], there is a KLJN line connecting the Certification Authority (CA) to the Roadside Devices (RSDs) and Roadside Key Providers (RSKP) (see Fig. 5).

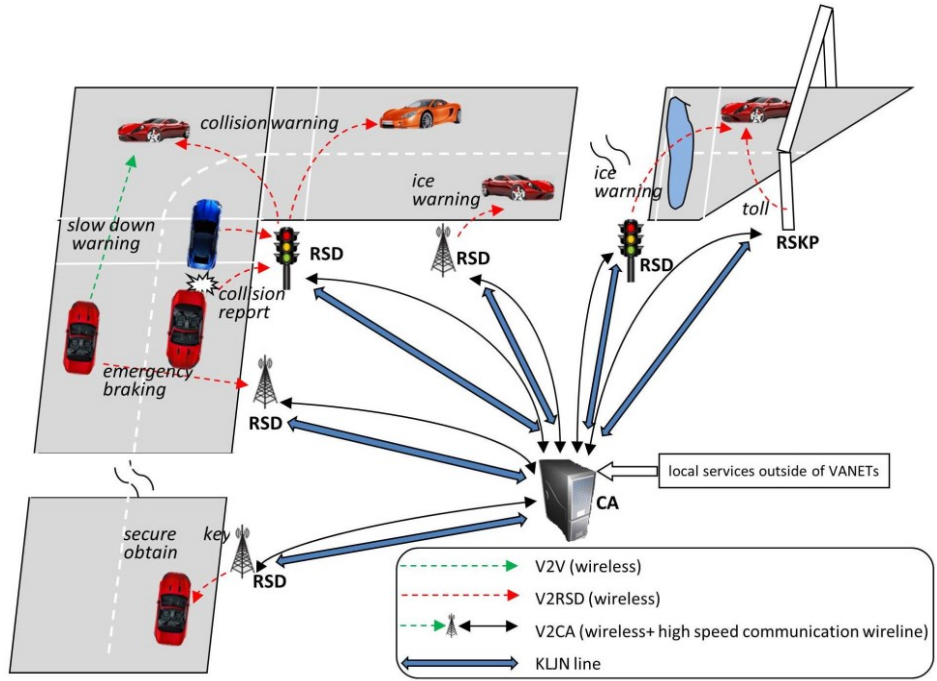

Fig. 5. Vehicular communication networks with unconditional secure key exchange. The network nodes remain the same except for a new node: the roadside key provider (RSKP) and extra wires for KLJN key exchange between the CA and RSD and/or RSKP. The existing wirelines between the RSDs/RSKPs and the CA are kept for high speed communication purposes.

The KLJN key generation process is performed as follows:
i). When a vehicle needs a secure key, it sends a message (via wireless communication) to the closest RSKP with the key request.
ii). The RSKP will use the extra wire (*i.e.,* the high speed communication line) to inform the CA in charge about the key request.
iii). A key generation process will take place between the RSKP and the CA.
iv). The RSKP will then provide the cars with the unconditional secure keys by using a near field communication wireless technology [9].
v). The RSDs also use their KLJN lines that connect them to the CA to generate KLJN keys that are used to secure the communication between RSDs and the CA.

Note that the KLJN line is used only to secretly generate and share the KLJN keys that are going to be used to secure the communication between two nodes. The rest of the communication is done either via wireless communication or using a high speed communication wireline.

### 2.2. *KLJN key donation in vehicular communication networks*

It is important to mention that the RSKP key donation that was proposed in [9], where RSKPs were visualized as gates, might not be as efficient as expected. This is because vehicles would have to slow down in order to get sufficiently close to the RSKPs (as proximity is needed for secure key donation). Therefore, we also propose a lane-by-lane key donation using RSKP equipment embedded in the pavement. In this way, vehicles will not have to slow down to obtain their keys. To detect vehicles in each lane, either loop detectors [29] or high-definition digital wave radars [30] deployed on the side of the roadway can be used. Both the RSKPs and the radar units can be connected to RSDs through a high speed wireline connection. Thus, the KLJN key generation is performed between RSDs and the CA only, while the RSKP will be only in charge of providing the cars with the unconditionally secure KLJN keys. Moreover, this key donation process would be encrypted with the former key, therefore, even if an eavesdropper is listening, he/she would not be able to extract the key information unless he/she has the former key. Figure 6 illustrates this solution.

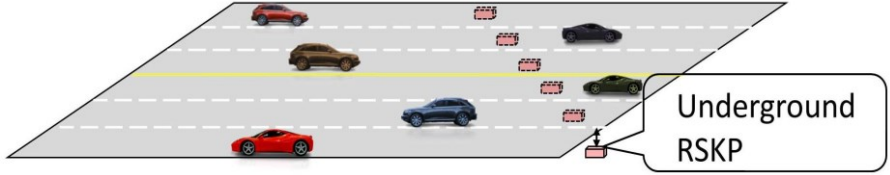

Fig. 6. Key donation with RSKP equipment embedded in the pavement. RSKPs are located underground of each lane.

### 2.3. *Upper limit of the KLJN key lifetime*

The lifetime of the KLJN key in vehicular communication networks is a very important technical parameter that needs to be discussed. This is because the longer the KLJN key is used, the more susceptible it is to attacks. In order to find out the lifetime of the KLJN key in vehicular communication networks, we proceed as follows.

First of all, the noise bandwidth $B_{KLJN}$ is determined by the distance $L$ between the two communicating parties, which in the case of vehicular communication networks depends on the length of the KLJN line segment between RSDs and the CA. Thus, the following relationship must be satisfied [19]: $B_{KLJN} \ll \frac{c}{L}$, where $c$ is the speed of electromagnetic waves in the wireline. Suppose that $0 < \Theta \ll 1$ and the noise bandwidth is:

$$B_{KLJN} = \Theta \frac{c}{L}. \tag{1}$$

Also, the duration of the bit sharing period $\tau$ must be long enough compared to the correlation time of the noise $\tau_{KLJN}$, i.e., $\tau_{KLJN} \approx \frac{1}{B_{KLJN}}$, in order to correctly distinguish between the different resistors situations [31, 32]. The frequency of *secure* bit exchange is:

$$f_{sec} = \frac{1}{2}\frac{B_{KLJN}}{\gamma}, \tag{2}$$

where $\gamma \gg 1$, see [31, 32] and the factor $\frac{1}{2}$ is due to the fact that a secure bit exchange occurs (on average) 50% of the time.

The lifetime of the KLJN key $\tau_k$ in vehicular communication networks depends on the vehicle density. For the sake of simplicity, first we assume homogenous car density:

$$n_c = \frac{N_c}{N_{KLJN}}, \tag{3}$$

where $N_c$ is the number of cars and $N_{KLJN}$ is the number of Roadside Devices with KLJN units. Thus, a KLJN unit serves $n_c$ cars. Consequently, the frequency of *secure* bit donation to a single car is:

$$f_c = \frac{f_{sec}}{n_c}. \tag{4}$$

If the length of the KLJN key is defined as $N_k$, then by combining Eqs. (1)–(4), we find that the lifetime of the KLJN key in vehicular communication networks is:

$$\tau_k = \frac{N_k}{f_c} = \frac{2N_k n_c \gamma L}{\Theta c}. \tag{5}$$

Note that this result represents a pessimistic estimation for inhomogeneous vehicular communication networks when $n_c$ is the upper limit of the number of cars any RSD is handling. Thus, Eq. (5) gives an upper limit of the lifetime of the KLJN key in vehicular communication networks. To demonstrate the results, we assign possible practical values to the parameters. Let $L = 1000\ m$, $c = 2*10^8\ m/s$, $\gamma = 100$ (since $\gamma = \frac{B_{KLJN}}{f_B}$, where $f_B = \frac{1}{\tau}$ should be low enough compared to $B_{KLJN}$, see [30,31]), $N_k = 100\ bits$, $n_c = 1000\ vehicles$, and $\Theta = 0.1$ (in order to satisfy $B_{KLJN} \ll \frac{c}{L}$, that is the "no-wave limit" condition [22]). Then the lifetime of KLJN key is $\tau_k = 10^3\ s$.

Techniques such as building parallel channels by using chip and multi-wire cables can be used to enhance the speed of the KLJN scheme and to decrease $\tau_k$ [19]. There is also a possibility to increase the security of physically exchanged keys in the case of repeated usage [33].

## 3. Conclusion

In this paper, we reviewed the communication infrastructure and discussed some security-related aspects of vehicular communication networks. We have proposed a KLJN key donation solution for vehicular communication networks. The KLJN key generation in vehicular communication networks has also been discussed and an upper limit for the lifetime of this KLJN key was computed.

**Acknowledgements**

X. Cao's contribution is supported by China Scholarship Council. Y. Saez is grateful to IFARHU/SENACYT for supporting her PhD studies at Texas A&M.